\documentclass[12pt]{iopart}
\usepackage{graphicx}
\usepackage{booktabs}

\begin{document}

\title[Anisotropic magnetic properties and CEF studies on CePd$_2$Ge$_2$ single crystal]{Anisotropic magnetic properties and crystal electric field studies on CePd$_2$Ge$_2$ single crystal }
\author{Arvind Maurya, R. Kulkarni, S. K. Dhar and A. Thamizhavel$^*$}
\address{Department of Condensed Matter Physics and Materials Science, Tata Institute of Fundamental Research, Homi Bhabha Road, Colaba, Mumbai 400 005, India}
\ead{$^*$thamizh@tifr.res.in}

\begin{abstract}

The anisotropic magnetic properties of the antiferromagnetic compound CePd$_2$Ge$_2$, crystallizing in the tetragonal crystal structure have been investigated in detail on a  single crystal grown by Czochralski method.  From the electrical transport, magnetization and heat capacity data,  the N\'{e}el temperature is confirmed to be 5.1~K.  Anisotropic behaviour of magnetization and resistivity is observed along the two principal crystallographic directions viz., [100] and [001].  The isothermal magnetization measured in the magnetically ordered state at 2~K exhibits a spin re-orientation at 13.5~T for field applied along [100] direction, whereas the magnetization was linear along the [001] direction attaining a value of 0.94~$\mu_{\rm B}$/Ce at 14~T. The reduced value of the magnetization is  attributed to the crystalline electric field (CEF) effects.  A sharp jump in the specific heat at the magnetic ordering temperature is observed.  After subtracting the phononic contribution, the jump in the heat capacity amounts to 12.5~J/K mol which is the expected value for a spin $\frac{1}{2}$ system.   From the  CEF analysis of the magnetization data  the excited crystal field split energy levels were estimated to be at 120~K and 230~K respectively, which quantitatively explain the observed Schottky anomaly in the heat capacity.  A magnetic phase diagram has been constructed based on the field dependence of magnetic susceptibility and the heat capacity data.  

\end{abstract}

        
\submitto{\JPCM}

\maketitle
\section {Introduction}

One of the most widely investigated topics in the field of condensed matter physics is the magnetism exhibited by Ce-based intermetallic compounds.  In some Ce compounds the $4f$ level lies in close proximity to the Fermi level,  enhancing the interaction between the conduction electrons and the $f$ electron leading to Kondo effect which screens the $4f$ derived magnetic moment.  The competition between the on-site Kondo effect with an energy scale $T_{\rm K}~\approx ~{\rm exp}(-1/[J N(E_{\rm F}])$ and the inter-site Ruderman-Kittel-Kasuya-Yosida (RKKY) magnetic interaction with an energy scale $T_{\rm RKKY}~\approx~[J^2 N(E_{\rm F})$], where $J$ is the exchange coupling between the local moment and the conduction electrons and $N(E_{\rm F})$ is the density of states at the Fermi level $E_{\rm F}$, leads to various diverse ground states like magnetic ordering, valence instability, heavy fermion nature, unconventional superconductivity etc.; hence Ce based intermetallic compounds have been widely investigated for several decades.  The CeT$_2$X$_2$ compounds, where $T$ is a transition metal and $X$ is Si or Ge, crystallizing in the very well known ThCr$_2$Si$_2$ - type body centered tetragonal crystal structure, have been particularly useful in this regard.  For example,  heavy-fermion superconductivity is observed in CeCu$_2$Si$_2$~\cite{Steglich}, pressure induced superconductivity in CePd$_2$Si$_2$~\cite{Sheikin}, CeRh$_2$Si$_2$~\cite{Movshovich, Araki}, CeCu$_2$Ge$_2$~\cite{Jaccard}, unconventional metamagnetism is observed in CeRu$_2$Si$_2$~\cite{Haen}. The pressure induced superconductor CePd$_2$Si$_2$ has been investigated in greater detail on a single crystal by Van Dijk~\textit{et al}~\cite{Dijk}.  From the aniostropic magnetic studies, they observed a change in the easy axis direction from $a$-axis to $c$-axis at around 50~K. The reduced moment and a relatively large Sommerfeld coefficient $\gamma$  in CePd$_2$Si$_2$ was attributed to Kondo effect with a strong spin fluctuation.  The neutron diffraction experiments both on polycrystalline sample and  single crystalline sample confirmed the antiferromagnetic ordering with the spins pointing along the [110] direction~\cite{Grier, Dijk}.  On the other hand, the magnetic properties of iso-electronic CePd$_2$Ge$_2$  have been investigated only on polycrystalline samples~\cite{Besnus, Feyerherm, Oomi, Fukuhara}.  Besnus \textit{et al}~\cite{Besnus}  reported the magnetic and thermal properties of CePd$_2$Ge$_2$ which orders antiferromagnetically at $T_{\rm N}$~=~5.1~K. They estimated the crystal field parameters from the magnetic susceptibility and from the Schottky heat capacity and inferred the crystal field  split energy levels to be located at  0, 110 and 220 K corresponding to the three doublets.  The magnetic structure of CePd$_2$Ge$_2$ was determined by Feyerherm~\textit{et al} and they found that the magnetic moment points along the [110] direction~\cite{Feyerherm} similar to the case of CePd$_2$Si$_2$.  Furthermore, they attributed  the reduced ordered moment of CePd$_2$Ge$_2$  not due to Kondo effect but due to the CEF and the anisotorpic exchange interaction~\cite{Feyerherm}.  The effect of chemical pressure by substituting Ni in place of Pd resulted in a decrease of $T_{\rm N}$~\cite{Fukuhara} whereas the hydrostatic pressure studies on resistivity measurement  revealed that the $T_{\rm N}$ increases with increasing pressure up to  values as high as  10~GPa~\cite{Oomi, Jaccard2}.  In continuation of our successful efforts in growing the single crystals and investigating the anisotropic magnetic properties of  CeT$_2$Ge$_2$ type compounds (where T is a transition metal) like CeAg$_2$Ge$_2$~\cite{Thamizh}, CeAu$_2$Ge$_2$~\cite{Devang} and CeCu$_2$Ge$_2$~\cite{Deepak},  here we report on the anisotropic magnetic properties of CePd$_2$Ge$_2$ single crystal by measuring the  transport and magnetic properties.  CEF analysis has been performed on the magnetic susceptibility and  heat capacity data to estimate the crystal field level splitting.

\section{Experiment}

The single crystal of CePd$_2$Ge$_2$ was grown by Czochralski crystal pulling method  in a tetra-arc furnace.  The starting materials of Ce (99.9\% purity), Pd (99.99\%) and Ge (99.9999\%) were taken in the stoichiometric ratio for a total mass of about 10-11 g.  The sample was remelted 4 times to ensure proper homogeneity in a water cooled copper hearth.  A polycrystalline tungsten rod was used as a seed and the crystal was pulled out of the melt at a rate of 10~mm/hr.   The pulled ingot was then subjected to powder x-ray diffraction (XRD),  using a PANalytical x-ray diffractometer with monochromatic Cu-K$_{\rm \alpha}$ radiation,  to check the phase purity. The orientation of the crystal was done by back reflection Laue  method.   The dc magnetic susceptibility and the magnetization measurements were performed in the temperature range 1.8-300~K using a superconducting quantum interference device (SQUID) and vibrating sample magnetometer (VSM).   The electrical resistivity was measured down to 1.9~K in a home made set up.  The heat capacity  was measured using a Quantum Design physical property measurement system (PPMS);  we have also used the dilution insert of Quantum Design to measure the heat capacity down to 50~mK and in applied magnetic fields as high as 14~T.

\section{Results}
\subsection{X-ray diffraction}

\begin{figure}[!]
\centering
\includegraphics[width=0.5\textwidth]{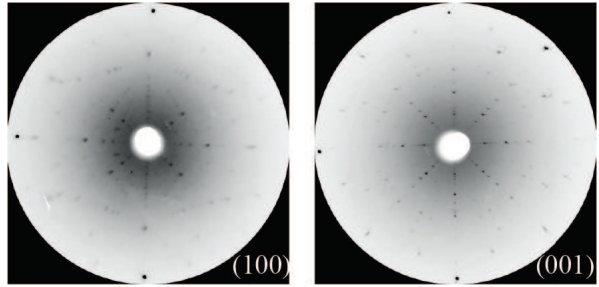}
\caption{\label{fig1}(Color online) Laue diffraction pattern of CePd$_2$Ge$_2$ single crystal corresponding to the (100) and (001) planes.}
\end{figure}

To confirm the phase purity of CePd$_2$Ge$_2$ a small portion of the crystal was subjected to powder x-ray diffraction.  The XRD  revealed  a clear pattern without any impurity peaks suggesting the phase purity of the grown crystal.  A Le-Beil fit was performed on the x-ray diffraction pattern and the lattice constants were estimated to be $a$~=~4.337~\AA~ and $c$~=~10.051~\AA~ which are in agreement with the previously reported values~\cite{Rossi}.   The composition of the crystal was further confirmed by energy dispersive analysis by x-ray (EDAX).   In order to study the anisotropic physical properties, the grown single crystal was cut along the principal crystallographic directions viz., [100] and [001].  This was accomplished by performing a back-reflection Laue.  The  Laue diffraction patterns corresponding to the (100) and (001) planes are shown in Fig.~\ref{fig1}.  Well defined Laue diffraction spots together with the tetragonal symmetry pattern confirmed the good quality of the grown crystal.  The crystal was then cut along the principal crystallographic direction using a spark erosion cutting machine for the measurement of resistivity, susceptibility and heat capacity.  

\subsection{Magnetization}

\begin{figure}
\centering
\includegraphics[width=0.45\textwidth]{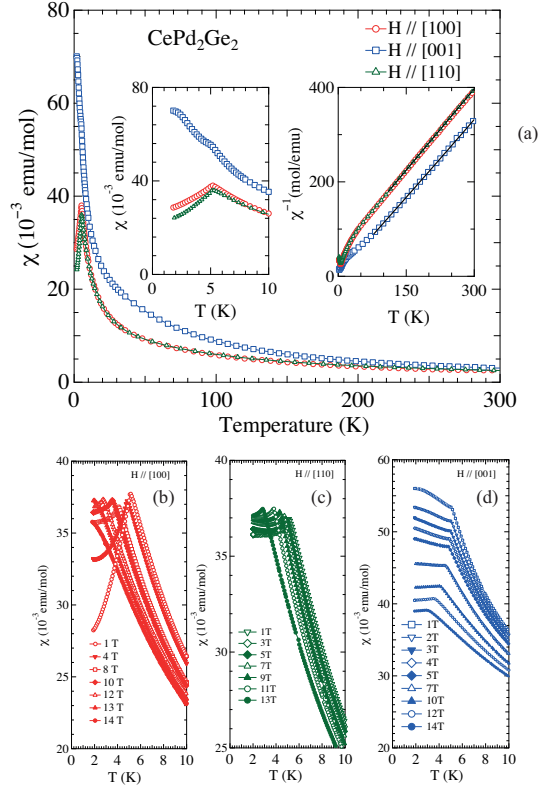}
\caption{\label{fig2}(Color online) (a) Temperature dependence of magnetic susceptibility of CePd$_2$Ge$_2$ for magnetic field parallel to [100], [110] and [001] directions.  The left inset shows the low temperature part where the magnetic ordering is seen clearly.  The right inset shows the inverse magnetic susceptibility  of CePd$_2$Ge$_2$, the solid lines are fit to the  Curie-Weiss law. (b), (c) \& (d) Low temperature part of the magnetic susceptibility in various applied magnetic fields along [100], [110] and [001] directions, respectively. }
\end{figure}

The temperature dependence of magnetic susceptibility of CePd$_2$Ge$_2$ is shown in the main panel of Fig.~\ref{fig2}(a), measured in an applied magnetic field of 0.1~T from 1.8~K to 300~K.  The magnetic susceptibility for $H~\parallel$~[001] direction is larger than for $H~\parallel$~[100], both in the paramagnetic and in the magnetically ordered state, reflecting the anisotropy.  For temperatures fairly higher than $T_{\rm N}$ the susceptibility shows Curie-Weiss like behaviour.  The low temperature part of the magnetic susceptibility along  the three crystallographic  directions [100], [110] and [001]  is shown in the left inset of Fig.~\ref{fig2}(a).  For field parallel to [100] direction, the susceptibility exhibits a clear cusp at 5.1~K indicating the antiferromagnetic transition at this temperature.  
On the other hand, the magnetic susceptibility shows a kink followed by an increase at 5.1~K for the [001] direction.  Similar behaviour is also observed for the isostructural CePd$_2$Si$_2$ compound~\cite{Dijk}.  Typically in a collinear two sublattice antiferromagnet, when the field is applied parallel to the moment direction, the susceptibility will apparently decrease to zero at $T~\rightarrow~0$ and for fields orthogonal to the moment direction the susceptibility will remain almost constant, below $T_{\rm N}$.  The increase in the susceptibility along [001] direction and a relatively large drop in the susceptibility below ($T_{\rm N}~=)$~5.1~K in the ab-plane suggests that [001] direction may be the hard axis of magnetization and the magnetic moment must be lying in the $ab$-plane as predicted by neutron diffraction experiments on polycrystalline samples~\cite{Feyerherm}.  From our susceptibility measurement it is obvious that the magnetic susceptibility falls more rapidly for $H~\parallel$~[100] direction, than along the other two directions, suggesting [100] is the easy axis of magnetization. The inverse magnetic susceptibility of CePd$_2$Ge$_2$ is shown in the right inset of Fig.~\ref{fig2}.  At high temperatures the inverse susceptibility is linear and at low temperature there is a deviation from this linearity.  This deviation is attributed to the crystalline electric field effect which is discussed later.  The high temperature part of the magnetic susceptibility is fitted to the  Curie-Weiss law, $\chi~= \frac{C}{T - \theta_{\rm p}}$, where $C$ is the Curie constant and $\theta_{\rm p}$ is the paramagnetic Weiss temperature.  We  obtain  $\theta_{\rm p}$ = $-47$~K and $\mu_{\rm eff}$ = 2.65~$\mu_{\rm B}$/Ce,  $\theta_{\rm p}$ = $-46$~K and $\mu_{\rm eff}$ = 2.64~$\mu_{\rm B}$/Ce  and  $\theta_{\rm p}$ = -5~K and $\mu_{\rm eff}$ = 2.74~$\mu_{\rm B}$/Ce  for $H~\parallel$~[100], [110] and [001] directions, respectively.  The negative sign of $\theta_{\rm p}$  suggests the antiferromangetic ordering and the effective magnetic moment is close to the free ion value of Ce$^{3+}$.  We  also measured the susceptibility  in various applied fields along the two principal crystallographic directions.  It is evident from Fig.~\ref{fig2}(b) that for fields parallel to [100] direction the antiferromagnetic transition shifts to lower temperatures.  For magnetic fields greater than 12.5~T, the susceptibility does not show any clear evidence of magnetic ordering down to 1.8~K. Apparently the magnetic transition become very broad at high fields around 14~T.  On the other hand a clear anomaly persists at low temperature pertaining to the magnetic ordering even at fields as high as 14~T along the [110] and [001] direction.  This supports the claim that [001]-axis is the hard axis of magnetization.  

\begin{figure}[!]
\centering
\includegraphics[width=0.45\textwidth]{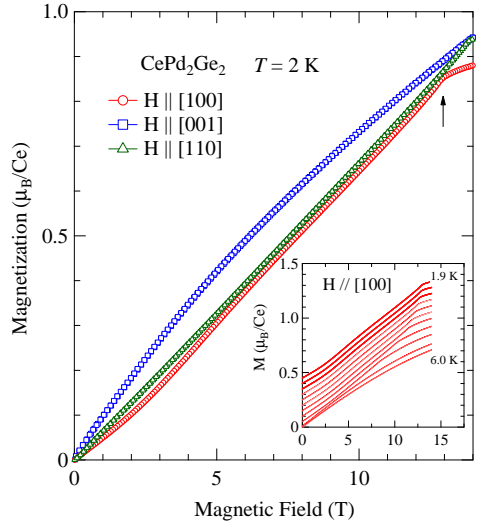}
\caption{\label{fig3}(Color online) Isothermal magnetizatin of CePd$_2$Ge$_2$ measured at $T~=$~2~K for $H~\parallel$~[100] and [001] directions. The inset shows the isothermal magnetization measured at various fixed temperatures for $H~\parallel$~[100].  The magnetization curves have been shifted up for clarity.  The  plots from the top are for temperatures 1.9, 2.2, 2.4, 2.6, 2.8, 3, 3.5, 4, 5 and 6~K.   }
\end{figure}

The anisotropic magnetic behaviour of CePd$_2$Ge$_2$ was further investigated by performing isothermal magnetization measurements $M(H)$ at a few selected temperatures.  The main panel of Fig.\ref{fig3} shows the magnetization  of CePd$_2$Ge$_2$ at 2~K along the two principal crystallographic directions. For $H~\parallel$~[100], the magnetization initially increases linearly with the field followed by a  small change in the slope at 2.4~T,  and then it increases almost lineary up to 13~T at which point there is another clear change of slope indicating the spin reorientation at these two fields.  On the other hand the magnetization for $H~\parallel$~[001] direction simply increases with the increase in field with a positive curvature and does not exhibit any evidence of spin re-orientation.  The magnetization along [110] direction almost overlaps with that of [100] direction, however the spin reorientation observed along [100] direction is not seen. The magnetization reaches a value of 0.88~$\mu_{\rm B}$/Ce and 0.94~$\mu_{\rm B}$/Ce at 14~T along [100] and [001] respectively.    These values of magnetization are much smaller than the theoretical value of $g_{\rm J}J$ (=$\frac{6}{7} \times \frac{5}{2}$), 2.14~$\mu_{\rm B}$/Ce.  The inset of Fig.~\ref{fig3} shows the $M(H)$ measured at various temperatures for $H~\parallel$~[100].  It is evident  that the spin reorientation occurring at 13.4~T at 2~K decreases to lower values as the temperature increases.  Close to the magnetic ordering temperature 5~K, the spin re-orientation is not discernible.

\subsection{Electrical Resistivity}

\begin{figure}[h]
\centering
\includegraphics[width=0.4\textwidth]{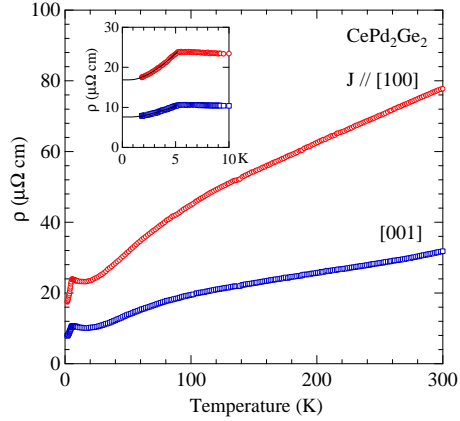}
\caption{\label{fig4}(Color online) Temperature dependence of electrical resistivity of CePd$_2$Ge$_2$  for current parallel to the two principal crystallographic directions.  The inset shows the low temperature part of the resistivity and the solid lines are the fits to antiferromagnetic spin-wave gap expression (see text).}
\end{figure}

Figure~\ref{fig4} shows the electrical resistivity of CePd$_2$Ge$_2$ measured in the temperature range from 1.8 to 300~K for current parallel to the two principal crystallographic directions, respectively.  There is a significant anisotropy in the electrical resistivity reflecting the tetragonal symmetry of the crystal and most likely an anisotropic Fermi surface.  The low temperature part of the electrical resistivity is shown in the inset of Fig.~\ref{fig4}.  The sudden drop in the electrical resistivity at 5.1~K is attributed to the  antiferromagnetic ordering, which occurs due to the reduction in the spin-disorder scattering.  As the sample is cooled below 300~K the resistivity decreases typical of a metallic sample.  The absolute value of electrical resistivity at 300~K is 77~$\mu \Omega \cdot$cm and 32~$\mu \Omega \cdot$cm for $J~\parallel$~[100] and [001] respectively which decrease to  17~$\mu \Omega \cdot$cm and 8~$\mu \Omega \cdot$cm at 1.8~K, respectively.  It is to be mentioned here that in CePd$_2$Si$_2$ also the electrical resistivity is larger along [100] direction than that in the [001] direction~\cite{Dijk}.  However, unlike the case of CePd$_2$Si$_2$, where  a clear $-{\rm ln}(T)$ behaviour is seen below 50~K, no such anomaly is observed in CePd$_2$Ge$_2$ this may be attributed to the larger unit cell volume of CePd$_2$Si$_2$, which does not favour Kondo effect.
A broad hump centered around 100~K is observed along both the directions which is attributed to the crystal field effect. Assuming that the excitation of spin waves in governed by the dispersion relation~\cite{Fontes}, 

\begin{equation}
\label{eqn1}
\epsilon_{\rm k} = \sqrt{\Delta^2 + D k^2},
\end{equation}

where $\epsilon_{\rm k}$ is the energy of the excitations, $\Delta$ is the gap in the spin-wave spectrum and $D$ is the spin-wave stiffness, the electrical resistivity in the ordered state is given by~\cite{Fontes},

\begin{equation}
\label{eqn2}
\rho(T) = \rho_{\rm 0} + \rho_{\rm AF} \Delta^2 \sqrt{\frac{T}{\Delta}}e^{-\frac{\Delta}{T}} \left[1 + \frac{2}{3}\left(\frac{T}{\Delta} \right)+\frac{2}{15}\left(\frac{T}{\Delta} \right)^2 \right],
\end{equation}

where the coefficient $\rho_{\rm AF}$ is proportional to $\frac{1}{D^{3/2}}$ and $\rho_{\rm 0}$ is the residual resistivity which is temperature independent. Eq.~\ref{eqn2} was fitted to the low temperature electrical resistivity data of CePd$_2$Ge$_2$ below T$_N$ and the obtained fitting parameters are $\rho_{\rm 0} = 16.873~\mu \Omega \cdot$cm, $\rho_{\rm AF} = 0.399~\mu \Omega \cdot$cm/K$^2$ and $\Delta = 3.962$~K for $J~\parallel$~[100] and $\rho_{\rm 0} = 7.592~\mu \Omega \cdot$cm, $\rho_{\rm AF} = 0.172~\mu \Omega \cdot$cm/K$^2$ and $\Delta = 3.814$~K for $J~\parallel$~[001]. Although the above equation is valid for $T~\ll~\Delta$, it provides a good fit to our experimental data in the magnetically ordered state right upto T$_N$. The spin-wave gap is almost the same for both the crystallographic directions and it is comparable to T$_N$ as observed in  other antiferromagnet Ce compounds~\cite{Pikul, Maria}.

\subsection{Heat Capacity}

\begin{figure}[h]
\centering
\includegraphics[width=0.4\textwidth]{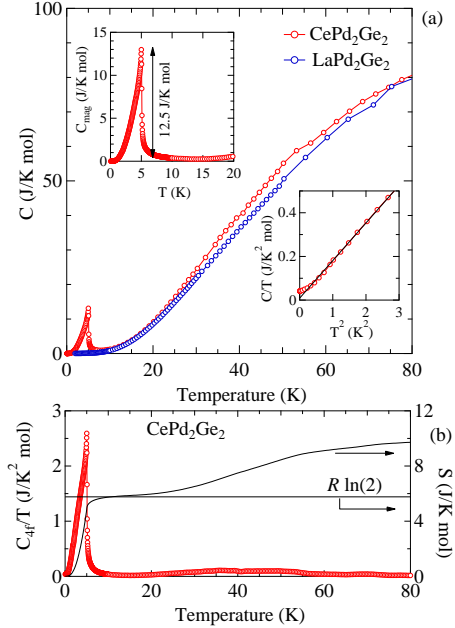}
\caption{\label{fig5}(Color online) (a) Temperature dependence of the specific heat capacity of CePd$_2$Ge$_2$ measured in the temperature range from 0.05 to 80~K  and LaPd$_2$Ge$_2$ measured in the temperature range from 1.8 to 80~K.  The top inset shows the magnetic part of heat capacity and the lower inset shows the $C/T$ vs. $T^2$ plot of CePd$_2$Ge$_2$. (b)  C$_{\rm mag}$/T vs. T of CePd$_2$Ge$_2$.  The calculated entropy is plotted in the right axis.}
\end{figure}

Figure ~\ref{fig5} shows the temperature dependence of the heat capacity $C_{\rm p}$ of CePd$_2$Ge$_2$ measured in the temperature range 0.05 to 80~K. Also shown in the main panel of Fig.~\ref{fig5} is the specific heat capacity of the non-magnetic analogue LaPd$_2$Ge$_2$ in the temperature range from 1.8 to 80~K.
The observed heat capacity of LaPd$_2$Ge$_2$ is typical for a non-magnetic reference compound.  The low temperature part of the heat capacity of LaPd$_2$Ge$_2$ was fitted to the expression $C/T = \gamma + \beta T^2$ to obtain the electronic specific heat coefficient $\gamma$ and the lattice contribution $\beta$.   The $\gamma$ and $\beta$ were estimated to be 8.03~mJ/K$^2 \cdot$mol and 0.427~mJ/K$^4$mol.  From the $\beta$ value one can estimate the Debye temperature $\Theta_{\rm D}$ as 283~K, which is typical of most of the La-based compounds.   The specific heat capacity  of CePd$_2$Ge$_2$ shows a  huge jump at 5.1~K thus confirming the bulk magnetic ordering.  The $C_{\rm p}$ of CePd$_2$Ge$_2$ is  higher than the non-magnetic reference compound in the temperature range investigated in the present work.  An estimate of the Sommerfeld coefficient $\gamma$ for CePd$_2$Ge$_2$ was obtained by the same method as employed for LaPd$_2$Ge$_2$ from the data below 1.7~K as shown in the lower inset of Fig.~\ref{fig5} .  The $\gamma$ value thus obtained is 9.17~mJ/K$^2$~mol which is close to that of LaPd$_2$Ge$_2$, thus indicating that the strength of the hybridization between the $4f$ electron and the conduction electrons in CePd$_2$Ge$_2$ is very weak.  The $4f$ contribution to the heat capacity $C_{\rm 4f}$ was obtained by subtracting the specific heat of LaPd$_2$Ge$_2$ from that of CePd$_2$Ge$_2$.  The top inset in Fig.~\ref{fig5}(a) shows the $C_{\rm 4f}$, where the jump in the heat capacity at $T_{\rm N}$ amounts to 12.5~J/K  mol.  In the mean field approximation the discontinuity in the  magnetic part of the heat capacity for a spin $S(=\frac{1}{2})$ system is given by~\cite{Buschow},
\begin{equation}
\label{eqn3}
\Delta C_{\rm mag}(T_{\rm N}) = 2.5 R \left[\frac{(2S + 1)^2-1}{(2S+1)^2+1}\right],
\end{equation}
where $R$ is the gas constant. The jump in the heat capacity observed in CePd$_2$Ge$_2$ exactly matches with the theoretical model.  The estimated magnetic entropy obtained by integrating $C_{\rm 4f}$/T is shown in Fig.~\ref{fig5}(b).  The entropy reaches 90\% of $R {\rm ln}(2)$ at the ordering temperature and recovers the full value for a doublet ($R{\rm ln(2)})$ at 10~K  thus indicating a doublet ground state well seperated from the first excited state.     Above $T_{\rm N}$ the entropy increases gradually and reaches $R$~ln~4  for temperatures greater than 100~K.  This roughly gives the estimate of the first excited state of the crystal field split level which is discussed in the next section.  
\begin{figure}[h]
\centering
\includegraphics[width=0.4\textwidth]{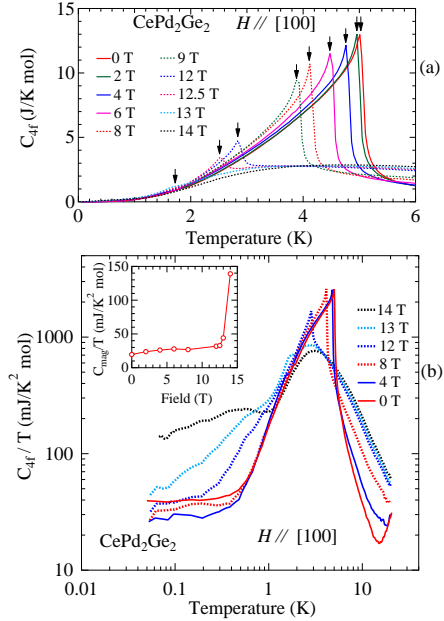}
\caption{\label{fig6}(Color) (a) Magnetic part of the heat capacity ($C_{\rm mag}(T)$) in the temperature range form 0.05 to 6~K of CePd$_2$Ge$_2$, in various applied magnetic fields along the [100] direction.  The arrows indicate the antiferromagnetic ordering.  (b)  Log-Log plot of $C_{\rm mag}/T$ of CePd$_2$Ge$_2$ in various applied magnetic fields in the temperature range from 0.05 to 20~K.  The inset shows the value of $C_{\rm mag}/T$ at 0.05~K.  }
\end{figure}

Figure~\ref{fig6} shows the magnetic part of the heat capacity measured in various applied magnetic fields, up to 14~T with $H~\parallel$~[100] direction.   It is evident  that the antiferromagnetic transition shifts to lower temperature, typical for an antiferromagnetic system. The shift to lower temperatures with the field corresponds well with the data depicted in Fig.~\ref{fig2}(b). The peak due to antiferromagnetic order is discernible up to fields as high as 13~T, where $T_{\rm N}$ has decreased down to 1.7~K.  For an applied field of 14~T the peak is not seen down to 0.05~K; instead a broad hump is observed.  From the log-log plot of $C_{\rm mag}/T$, one can see from Fig.~\ref{fig6}(b) that for fields greater than 12~T, the curves fall almost on a single trace in the paramagnetic region. The inset of Fig.~\ref{fig6} shows that $C_{\rm mag}$/T value close to 0.05~K increases rapidly  beyond 13 T. Similar type of behaviour is seen in CeAuSb$_2$ (T$_N$=6~K) which exhibits field induced quantum critical point at 5.4~T and $C_{\rm mag}$/T increases rapidly in the vicinity of critical point and shows a peak at the critical field~\cite{Balicas}.  
  
\section{Discussion}

From the magnetic susceptibility, electrical resistivity and heat capacity measurements it is inferred that the single crystal of  CePd$_2$Ge$_2$ undergoes an antiferromagnetic ordering at $T_{\rm N}$ = 5.1~K which is coincident with the $T_{\rm N}$ of polycrystalline samples reported by Besnus \textit{et al}.~\cite{Besnus, Oomi}.  The field dependence of magnetic susceptibility and the heat capacity on single
\begin{figure}[h]
\centering
\includegraphics[width=0.4\textwidth]{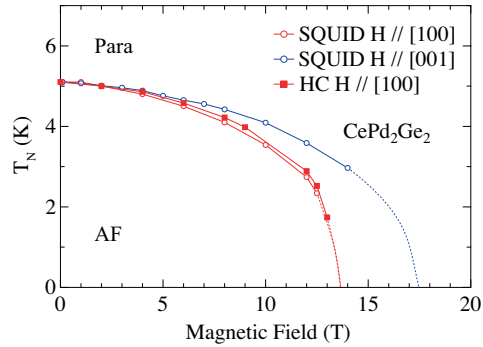}
\caption{\label{fig7}(Color online) Magnetic phase diagram of CePd$_2$Ge$_2$.  The dashed lines are guide to the eyes.  }
\end{figure}
crystalline sample enabled us to construct a magnetic phase diagram which is shown in Fig.~\ref{fig7}.   The N\'{e}el temperature above 6 T decreases much faster along [100] than that of [110] and [001] directions.  This suggests that [001] direction is the hard axis of magnetization.    Furthermore, the magnetization reaches only 0.88~$\mu_{\rm B}$/Ce and 0.94~$\mu_{\rm B}$/Ce for $H~\parallel$~[100] and [001] directions, respectively at 14~T.  These values are consistent with the neutron diffraction results on a polycrystalline sample where the ordered moment was reported to be 0.85~$\mu_{\rm B}$/Ce at 1.8~K~\cite{Feyerherm}. It is to be mentioned here that in the isostructural CePd$_2$Si$_2$, the ordered moment is only 0.62~$\mu_{\rm B}$/Ce.  This reduced value of moment in CePd$_2$Si$_2$ is attributed to the Kondo effect, where a $-{\rm ln}(T)$ behaviour is observed in the electrical resistivity and the jump in the heat capacity at the magnetic transition is lower, with the magnetic entropy amounting to $0.6 - 0.75~R{\rm ln}2$~\cite{Dhar, Sheikin2}.  Since no such behaviour is observed in CePd$_2$Ge$_2$ the reduced value of magnetization is mainly attributed to the crystal electric field effect.  

\begin{figure}[!]
\centering
\includegraphics[width=0.4\textwidth]{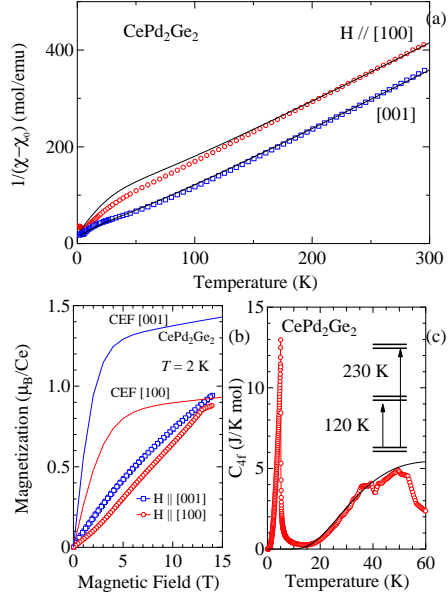}
\caption{\label{fig8} (Color online)  (a) Inverse magnetic susceptibility of CePd$_2$Ge$_2$.  The solid line is based on the CEF calculation, (b) The isothermal magnetization data of CePd$_2$Ge$_2$ along with the magnetization calculated based on the CEF parameters (c) magnetic part of the specific heat capacity of CePd$_2$Ge$_2$.  The solid line is calculated Schottky heat capacity. The obtained energy levels are also shown.}
\end{figure}

We analysed the magnetic susceptibility and the heat capacity  using the point charge CEF model.  The Ce atoms in CePd$_2$Ge$_2$ occupy the $2a$ Wyckoff's position and possesses the $\mathcal{D}_{\rm 4h}$ tetragonal point symmetry.  For tetragonal site symmetry,  for $J = 5/2$, the $2J+1$  6-fold degenerate  level splits into three doublets.   To understand the observed anisotropy in the magnetic susceptibility and to know  the crystal field energy splitting, we have performed the crystal field analysis on these data.  For this purpose in Fig.~\ref{fig8}(a) we have plotted the susceptibility data as $1/(\chi-\chi_{\rm 0})$, where $\chi_{\rm 0}$ was estimated as $-5.571~\times$~10$^{-5}$ and $1.230~\times$~10$^{-4}$~emu/mol for $H~\parallel$~[100] and [001] directions, respectively, so an effective magnetic moment of 2.54~$\mu_{\rm B}$/Ce is obtained for temperature above 50~K.  A similar  approach was taken for CePt$_3$Si and CeAg$_2$Ge$_2$ while performing the CEF analysis of the susceptibility data~\cite{Takeuchi, Thamizhavel}.  The CEF Hamiltonian for the tetragonal site symmetry  is given by

\begin{equation}
\label{eqn4}
\mathcal{H}_{{\rm CEF}}=B_{2}^{0}O_{2}^{0}+B_{4}^{0}O_{4}^{0}+B_{4}^{4}O_{4}^{4}+  B_{6}^{0}O_{6}^{0} + B_{6}^{4}O_{6}^{4},
\end{equation}
where $B_{\ell}^{m}$ and $O_{\ell}^{m}$ are the CEF parameters and
the Stevens operators, respectively~\cite{Hutchings,Stevens}. For Ce atom the $O_6$ Stevens operators are zero and hence the last two terms in the Hamiltonian are zero.

The magnetic susceptibility including the molecular field contribution $\lambda_{i}$ is given
by
\begin{equation}
\label{eqn5}
\chi^{-1}_{i} = \chi_{{\rm CEF}i}^{-1} - \lambda_{i}.
\end{equation}

The expression for the magnetic susceptibility based on the CEF model is given in our previous report~\cite{Pranab}.  We have calculated the inverse magnetic susceptibility for field along the two principal crystallographic directions based on the above CEF model.  
By diagonalizing the CEF Hamiltonian the eigenvalues and eigenfunctions are obtained.  The eigenvalues corresponds to the energy levels.  The solid lines in  Fig.~\ref{fig8}(a) are the calculated susceptibility with the unique values of the crystal field parameters given in Table~\ref{table1}.  The negative value of the 
\begin{table}
\begin{center}
\caption{\label{table1} CEF parameters, energy level schemes
and the corresponding wave functions for CePd$_2$Ge$_2$.}
\begin{tabular}{lccccccc}\hline
& & \multicolumn{2}{c}{CEF parameters} & \\ \hline
& $B_{2}^{0}$~(K) & $B_{4}^{0}$~(K) & $B_{4}^{4}$~(K) & $\lambda_{i}$~(emu/mol)$^{-1}$ \\
& $-4.18$ & $-0.25$ & $-3.75$ & $\lambda_{[100]}$ = $-17$ \\
&      &         &   & $\lambda_{[001]}$ = $-12$  \\ \hline
\multicolumn{2}{c}{energy levels and wave functions} &         &   &   \\ \hline
$E$(K) & $\mid+5/2\rangle$ & $\mid+3/2\rangle$ & $\mid+1/2\rangle$ & $\mid-1/2\rangle$ & $\mid-3/2\rangle$ & $\mid-5/2\rangle$ \\
230 & $0.8603$ & 0 & 0 & 0 & $0.5098$ & 0 \\
230 & 0 & $0.5098$ & 0 & 0 & 0 & $0.8603$ \\
120 & 0 & 0 & 1 & 0 & 0 & 0 \\
120 & 0 & 0 & 0 & 1 & 0 & 0 \\
0   & 0 &  $0.8603$ & 0 & 0 & 0 & $-0.5098$ \\
0   & $-0.5098$ & 0 & 0 & 0 & $0.8603$ & 0 \\
\hline
\end{tabular}
\end{center}
\end{table}
exchange field constant indicates the antiferromagnetic interaction between Ce moments. From the obtained eigenfunctions it is observed that there is a mixing of $\vert \pm \frac{5}{2} \rangle$  and $\vert \pm \frac{3}{2} \rangle$ wave functions in the ground and second excited state while  the first excited state is composed mainly of $\vert \pm \frac{1}{2} \rangle$.  These crystal field parameters reproduce the experimental susceptibility reasonably well and are in close agreement with the parameters estimated by Besnus~\textit{et al}~\cite{Besnus} from  the data obtained for a polycrystalline sample.  The obtained energy levels are $\Delta_{\rm 1} = 120$~K and $\Delta_{\rm 2} = 230$~K.  Figure~\ref{fig8}(b) shows the isothermal magnetization plot of CePd$_2$Ge$_2$ measured at $T~=~2$~K together with the calculated magnetization curves.  Although the CEF calculated magnetization curves do not reproduce the experimental data, it should be noted here that the anisotropy in the magnetization along the [100] and [001] directions are clearly explained.   
CePd$_2$Ge$_2$ is one of the systems where the magnetization, in the ordered state, is larger for the axis which is perpendicular to the direction of the moment orientation~\cite{Feyerherm}.  Similar kind of behaviour is seen in CeAgSb$_2$ where the magnetization increases very rapidly along the hard axis direction [100], than along the easy axis direction [001]~\cite{Takeuchi2} and the ordered moment points along the [001] direction~\cite{Araki2}.  A detailed neutron diffraction study on single crystal will lead to a more comprehensive understanding of the magnetic structure of CePd$_2$Ge$_2$.  

The magnetic part of the heat capacity of CePd$_2$Ge$_2$ obtained after subtracting the heat capacity of LaPd$_2$Ge$_2$, shown in Fig.~\ref{fig8}(c) exhibits a broad peak above 40~K which is attributed to the Schottky type excitations between the CEF levels of the Ce$^{3+}$ ions.  The Schottky heat capacity for a 3 level system (a ground state and two excited states) is given by the following expression
\begin{equation}
\label{eqn6}
\fl
C_{\rm Sch}= \left[\frac{R}{(k_{\rm B}T)^2} \frac{e^{(\Delta_1 + \Delta_2)/k_{\rm B}T}[-2 \Delta_1 \Delta_2 + \Delta_2^2 (1 + e^{\Delta_1/k_{\rm B}T}) + \Delta_1^2 (1 + e^{\Delta_2/k_{\rm B}T})]}{(e^{\Delta_1/k_{\rm B}T} + e^{\Delta_2 / k_{\rm B}T} + e^{(\Delta_1 + \Delta_2)/k_{\rm B}T})^2} \right]
\end{equation}
where $R$ is the gas constant and $\Delta_1$ and $\Delta_2$ are the crystal field split excited energy levels.   Having found the energy spacing for the first and the second excited doublets from the magnetic susceptibility, we have used the same energy values $\Delta_1$~=~120~K and $\Delta_2$~=~230~K in Eqn.~\ref{eqn6} and found that it explains the observed Schottky anomaly.  The solid line in Fig.~\ref{fig8}(c) shows the calculated Schottky heat capacity, thus validating our crystal field splitting estimation from the susceptibility data.  The discrepacy above 50~K ma be attributed to the difference in the phonon spectra of LaPd$_2$Ge$_2$ and CePd$_2$Ge$_2$ at higher temperatures.

The application of magnetic field shifts the antiferromagnetic ordering to lower temperatures which is more prominent along the [100] direction than along the [001] direction, as evident from the magnetic phase diagram shown in Fig.~\ref{fig7}.  The low value of the Sommerfeld coefficient $\gamma$ (9.2~mJ/K$^2 \cdot$mol) in CePd$_2$Ge$_2$  which is comprable to that of LaPd$_2$Ge$_2$ indicates that the hybridization between the conduction electron and the $f$ electron is weak and the $f$ electrons are highly localized here.  Furthermore, it is evident from the inset of Fig.~\ref{fig6}(b) that the $C_{\rm mag}/T$  increases to a large value in the magnetic field when the antiferromagnetic ordering vanishes, and is attributed to the spin fluctuations.    The enhancement in the $C_{\rm mag}/T$ value, when the antiferromagnetic ordering vanishes due to the application of magnetic field is also observed in CeAuSb$_2$, which shows a field induced quantum critical point~\cite{Balicas}.  It will be interesting to see the effect of magnetic field on the electrical resistivity of CePd$_2$Ge$_2$, down to very low temperature, which is planned for the future.  

\section{Conclusion}

The anisotropic magnetic properties of CePd$_2$Ge$_2$ have been investigated by growing a single crystal, in at tetra-arc furnace.  The phase purity and the crystal composition was confirmed by x-ray diffraction and EDAX.  The transport and magnetic properties have revealed large anistropy along the two principal crystallographic directions viz., [100] and [001], reflecting the tetragonal crystal structure.  The antiferromagnetic order is confirmed as $T_{\rm N}$ = 5.1~K.  The electrical resistivity in the ordered state can be well explained by the spin-wave gap model.  The N\'{e}el temperature was found to decrease with increasing field, as expected for a typical antiferromagnet system.  For fields greater than 13~T, the magnetic ordering was found to vanish.  Our crystal field calculation clearly explain the anisotropy in the magnetic susceptibility and the magnetization and the energy level thus obtained for the first and second excited state was found to be 120~K and 230~K.  These energy levels clearly explained the Schottky anomaly in the magnetic part of the heat capacity.  The jump in the magnetic part of the heat capacity was found to 12.5~J/K mol as expected by the mean field model for a spin 1/2 system.  There was no signature of Kondo effect and the magnetic ordering in this system is believed to be purely Ruderman-Kittel-Kasuya-Yosida (RKKY) type interaction.  One of the interesting findings is that although the magnetization is larger along the [001] direction as observed experimentally and supported by the CEF calculation, the moment direction is orthogonal to it, oriented in the $ab$-plane.

\end{document}